\documentstyle[epsfig,preprint,prl,aps]{revtex}
\begin{document}

\draft
\title{Bound states of $^3$He at the edge of a
$^4$He drop on a cesium surface}
\author{ R. Mayol,$^1$
 M. Barranco,$^1$
E.S. Hern\'andez,$^{2}$
M. Pi,$^1$   and M. Guilleumas$^1$} 
\address{$^1$Departament ECM,
Facultat de F\'{\i}sica, Universitat de Barcelona, E-08028 Barcelona,
Spain
}
\address{$^2$Departamento de F\'{\i}sica, Facultad de Ciencias Exactas y Naturales,
Universidad de Buenos Aires, 1428 Buenos Aires, Argentina}
\date{\today}

\maketitle

\begin{abstract}

We show that  small amounts of $^3$He atoms, added to a
$^4$He drop deposited on a flat cesium surface at zero temperature,
populate  bound states localized at the contact line. These edge states
 show up for drops large enough to develop  well defined surface
and bulk regions together with a contact line, and they 
 are structurally different from the well-known Andreev states that
appear at the free surface and at the liquid-solid interface of
films. We illustrate the  one-body density of $^3$He  in a drop with
1000  $^4$He atoms, and show that for sufficiently large number of
impurities,  the  density profiles spread beyond the edge, 
coating both the curved drop surface and its flat base and 
eventually isolating it from the substrate.

\end{abstract}


During the last  decade,  experimental and theoretical
studies of wetting phenomena on substrates of various
adsorptive strenghts have considerably enriched the physics of quantum
fluids.  Liquid  $^4$He, which  was early believed  to be  a universal
wetting agent,  exhibits finite  wetting temperatures  on cesium 
surfaces\cite{nacher91}. On the other hand,  the theoretical 
prediction that $^3$He is a  universal
 wetting agent\cite{pricaupenko94a} was confirmed by measurements
of its adsorption isotherms   on Cs\cite{ross97}. These opposite and
 complementary wetting  properties combine in a  very interesting  manner
 when  the helium  isotopes  coexist in  a low-temperature solution, since  $^3$He
dissolved into bulk liquid  or $^4$He films  populates a two-dimensional
homogeneous layer of Andreev states on  the free surface (see  e.g., Ref.
\onlinecite{hallock00} and refs. therein).

Andreev-like states for a single $^3$He atom, localized at the
liquid-substrate   interface of a $^4$He film on a variety of adsorbers, 
including weak alkalis like Cs, were predicted within the frame of density 
functional (DF) theory\cite{pavloff91b}.  The effect of $^3$He 
  impurities on the liquid-vapor and liquid-solid 
surface tensions  of $^4$He on Cs is a key ingredient in the prediction
of reentrant  wetting by  helium  mixtures\cite{pettersen93},  as  measured  shortly
afterwards\cite{ketola93,ross96}. The existence of the interfacial
bound states was 
confirmed by experimental  evidence\cite{ross95} and by microscopic variational
calculations\cite{clements97}. More recent measurements of the contact
angle of  helium mixtures on Cs\cite{klier99}   showed that  the large
observed values  are consistent  with both   single-particle (sp)
states of the  $^3$He atoms at the liquid-solid interface and surface
excitations.

The preference of $^3$He to  migrate to the surface of  $^4$He
is a  very general consequence  of the mutual interaction  between the
isotopes,  and this structure  appears as well in density profiles of
mixed   drops as described by DF
theory\cite{dalfovo89,barranco97,pi99}.  These  descriptions of finite helium mixtures 
 consider   freely standing systems in vacuum; however, mixed clusters  
deposited on adsorbing substrates offer an interesting illustration  of the
competition between selfsaturation, mutual interaction between the helium
isotopes and external confinement, with consequences on their wetting
behavior. A theoretical approach has been anticipated  for  droplets of 
pure $^4$He  on  Cs, whose profiles and contact angles have been recently   
computed\cite{ancilotto98}, employing a Finite
Range Density Functional  (FRDF)\cite{dalfovo95} at zero temperature.

In this letter, we present DF calculations of nanoscale mixed
$^3$He-$^4$He droplets at zero temperature on a flat Cs
surface using the FRDF for mixtures
of Ref.\onlinecite{mayol01} and demonstrate that the lowest-lying
$^3$He sp  bound states are localized 
on the perimeter of the  $^4$He base -i.e., the contact line.  These
are genuine
edge states of a different nature than the well-known Andreev ones, since
they originate in the peculiarities of the
mean field for this specific geometry.

For this sake, assuming axial symmetry around the $z$-axis 
perpendicular to the Cs substrate, by functional differentiation of the
energy density
we derive the Euler-Lagrange (EL) equation for the
$^4$He particle
density $\rho_4(r, z)$  and the Kohn-Sham (KS) equations for the
$^3$He  sp wave functions (wf's) 
$\psi_{nl}(r, z) \,e^{\displaystyle i l \varphi}$,  with $l$ the 
$z$-projection of the orbital angular momentum, for fixed particle
numbers $N_i$ ($i$=3, 4). The general  form of these
equations can  be found in Ref.\onlinecite{barranco97}; hereafter we
shall  mostly restrict our  considerations to the case 
$N_4$ = 1000 and varying
$N_3$  as discussed below. 

The first problem that we solve is the Schr\"odinger equation for one
$^3$He atom in the field of a $^4$He cluster on a Cs surface, with 
the Cs-He potential of Ref. \cite{ancilotto98}. In 
Fig. \ref{fig1} we show  contour plots on the $y=0$ plane
(hereafter, lengths are given in $\rm \AA$) 
of the probability densities 
$|\psi_{n0}(r, z)|^2$ for $n$ = 1, 2, and 3, together with grayscale views of 
the density
$\rho_4(r, z)$, which exhibits the layered structure encountered in  films 
on flat surfaces\cite{krotscheck85} (see also Ref. \cite{ancilotto98}).

The most spectacular result is undoubtedly the appearance of
the $1s$ ground state (gs) $\psi_{10}$, since the sharp 
localization of the impurity on the contact line  
indicates the existence of a new kind
of sp state, the edge state. 
With a truly one-dimensional (1D) probability density on
a ring of mesoscopic size,  it would  certainly  exist in a 
macroscopic
droplet, being in essence a 1D analog of the two-dimensional  
Andreev state at the planar free surface of $^4$He. Its
origin is the density distribution of the $^4$He drop, that together with
the potential created by the substrate, gives rise to the strong attraction
experienced by the impurity near the contact line. In addition, 
the enhanced surface
width  of the $^4$He cluster at the edge, due to the overlapping 
spreads of both the upper and the lower  surface, 
permits large zero point motion of the impurity, thus
lowering  its kinetic energy and favouring binding. 

We display in
Fig. \ref{fig2} the evolution of the sp energy $\varepsilon_{1s}$
with $N_4$, which shows that for relatively small sizes (around $N_4$=2000),
the curve seems to be asymptotic to a certain value, the macroscopic
limit. We have verified as well that $s$-states with
$n$ between 2 and 8  span an 
energy shell, with a width of about 1.3 K 
starting at $\varepsilon_{20}$= -4.74 K,
for i.e., $N_4$ = 1000 (cf. Fig. \ref{fig3}), that
could be regarded as fragmentation of the Andreev state of the free 
surface of films, induced by finite size and  curved shape. This
is supported  by the shape of the excited 
probability densities  displayed in Fig. \ref{fig1},
which display peaks on the upper surface of the drop.

Noticeably also, we have found that the  probability densities 
$|\psi_{1l}(r, z)|^2$  for $l >0$ are essentially identical to 
$|\psi_{10}(r, z)|^2$. 
This reflects the fact that these excitations consist of
 rotations of the  gs $\psi_{10}$ around 
the symmetry axis of the $^4$He drop, as can be seen in the sp level scheme
  depicted in Fig. \ref{fig3}.
Instead, since  $s$-excited states possess nonvanishing probability density
 on the rotational axis, states with $n>1$ and $l>0$
are more sensitive to the 
centrifugal potential, that splits the cusp  creating extra lobes in the 
higher levels described by wf's $\psi_{nl}(r, z)$.  

For the rotational band,
 the sp energies follow the rule
\begin{equation}
\varepsilon_{1l}= \varepsilon_{10} + \frac{\hbar^2 l^2}
{2 m^*_{10} R^2_{10}}
\label{eq1}
\end{equation}
with $\varepsilon_{10}$ = -5.03 K. Effective masses $m^*_{nl}$, defined
as the state averages of the local, density-dependent
one\cite{pavloff91b}, 
\begin{equation}
\frac{1}{m^*_{nl}} = \int d\;\vec{r}\, \frac{1}{m^*_3\left[\rho_4(r,
z)\right]} \,[\psi_{nl}(r, z)]^2
\label{eq2}
\end{equation}
 are parameters that appear naturally in the
prefactor of the centrifugal potential as one computes the expectation
value of the sp hamiltonian.  In this case, we obtain $m^*_{10}$ = 1.21 $m_3$,
which together with the slope of the linear regression in Fig. \ref{fig3}
 gives a gs radius $R_{10}$ = 35.1 $\rm \AA$. We point out that if
$m^*_{10}=m_3$, the corresponding values 
$\varepsilon_{10}$ = -4.86 K and $R_{10}$ = 34.7 $\rm \AA$ reveal
 the scarce relevance  of this parameter to our main results.
 For the  $s$-states $n$=2 to 8, the respective  $m^*_{n0}$ 
are similar and close to 1.3 $m_3$, that scales  with the 
 effective mass for the Andreev surface   state 
in $^4$He films on Cs\cite{pettersen93,clements97,treiner95},
around 1.45 $m_3$.

 As a second step, we have solved the system consisting of the coupled EL/KS
 equations, for  $N_3$ values corresponding to closed shell configurations
made of sp states
within the gs rotational band. For small $N_3$ values, the 
level filling sequence still corresponds to
the sp spectrum of a single atom, with slightly decreasing
 slopes of the gs rotational band,  due to the finite $^3$He density 
which enhances the effective mass, and  smaller values of
$|\varepsilon_{10}|$  also depending on $N_3$. This is illustrated 
in Fig. \ref{fig3}
for $N_3$ = 26.
 The shape of the wf's $\psi_{1l}(r, z)$ below the Fermi sea
is now sensitive to $N_3$, and $\rho_4(r,z)$ is also affected;
  maps of the $^3$He  and $^4$He 
 particle densities, with details  as in 
Fig. \ref{fig1}, are depicted in  Fig. \ref{fig4}, showing the
tendency of the $^3$He density to drift upwards on the drop surface; 
note that the
lowest $^3$He density line corresponds to 10$^{-6} \rm \AA^{-3}$.

A substantial  simplification of the above full
coupled problem,
within excellent accuracy for  sufficiently large $N_3$, is provided by
the Thomas-Fermi (TF) method\cite{stringari87}.
We apply
 this procedure to $N_3 \ge 100$, and find
 that at $N_3=100$, the density
 profile of the impurities covers the free
surface of the $^4$He drop, and tends to isolate it
from the substrate.  
For  $N_3$  between 100 and 150,
the situation qualitatively changes, and $^3$He spreads
beyond the edge of the $^4$He drop giving rise to a thin
film of $^3$He that coats the Cs substrate with an
asymptotic coverage $n_3$. As our interest here is in finite
systems, we have just
solved the problem of the splashed $^4$He$_{1000}$ droplet
for two very low asymptotic coverages, namely
$n_3 = 2 \times 10^{-3}$  and   $3 \times 10^{-3} {\rm \AA}^{-2}$.

 Figure \ref{fig5} is a  threedimensional view of $\rho_3(r,z)$
corresponding to 
$n_3 = 3 \times 10^{-3} {\rm \AA}^{-2}$,
where the occupied edge states of  the single atoms 
can be visualized as large peaks near the contact line,
while Fig. \ref{fig6} exhibits the corresponding  $\rho_4(r,z)$.
The layering of the $^3$He density
 appears as a series of peaks on the covering shell on the $^4$He
cluster; in addition, we note the spreading of $\rho_3(r,z)$ 
near the substrate as
$r$ increases beyond the region populated by the edge states. 
In this case,
the number of $^3$He atoms inside a cylinder of radius $R=50 {\rm \AA}$
is $\sim 180$, and $\sim 138$ for $n_3 = 2 \times 10^{-3} \, {\rm \AA}^{-2}$.

Finally, we comment on the contact angle, whose
determination  
for nanoscopic droplets
is far from trivial, as discussed in
\onlinecite{ancilotto98}. We have seen that helium droplet
 densities display 
non-negligible surface widths and are highly stratified near the
substrate.  This structure renders the contact angle 
an ill defined quantity, and one has to resort to
methods such as that in \onlinecite{ancilotto98}, since 
a determination of the contact angle by 
`visual inspection' of the equidensity lines may only lead to a
rather crude estimate.
For the pure $^4$He droplets our results coincide with those of
\onlinecite{ancilotto98}, thus  the procedure there described 
would give a contact angle of about 30 degrees.

The presence of $^3$He affects the contact
angle, as can be judged from Fig. \ref{fig7}, where we show the 
contour plot of the total $^3$He+$^4$He density corresponding to 
roughly half  the bulk  $^4$He density
for  $N_4=1000$ and the above two $^3$He coverages.
It appears from this figure that whereas the height of the cap does not 
change visibly, the radius of the cap base increases by $8-10 \%$
and this should have a sizeable effect in any sensible
determination of the contact angle.

 To summarize, we have found that in addition to the well known
 Andreev states, $^4$He drops on Cs substrates
host a new class of $^3$He sp states which appear as  1D 
rings of $^3$He atoms surrounding the contact line
for mesoscopic drops or wedges, and are the
lowest-lying. The gs energies found in this work are, for the largest clusters,
close to those of surface\cite{pavloff91b,treiner95}
Andreev states, around -5.1 K depending on the $^4$He coverage, and below
the interfacial\cite{ross95,treiner95} Andreev states, around  -4.3 K.
 The  experimental
detection of these edge states might be a challenge, open to seeking 
 thermodynamic evidence
through specific heat or magnetic susceptibility measurements below
 the dewetting line of the low temperature phase diagram for 
mixtures\cite{ross96}, or dynamical evidence for undamped 1D spin-density
 oscillations\cite{esh02}.
The statistical model  in Ref. \onlinecite{klier99} 
 shows that the most likely
scenario for measured contact angles and  surface tensions
 must include both Andreev states and ripplons at the substrate-liquid
interface; accordingly, it might be interesting to examine the effects
of the new edge states -so far
overlooked- in this description. 

This work has been performed under grants BFM2002-01868  from
DGI, and 2001SGR-00064 and ACI2000-28
from Generalitat of Catalunya.
E.S.H. has been also funded by M.E.C.D. (Spain)
on sabbatical leave.  We are grateful
to Francesco Ancilotto, Milton Cole and
Adrian Wyatt for useful discussions, and to Eckhard Krotscheck for helpful advice.

\begin{figure}
\caption[]{Probability densities of   $s$-states
of a $^3$He atom together with the $^4$He particle density
 in the $y=0$ plane for a drop with $N_4$=1000.}
\label{fig1}
\end{figure}

\begin{figure}
\caption[]
{Energy of the   $1s$ state 
for $N_4=20$, 100, 500, 1000, 1500, 2000, and 3000.
The calculated energy of one single $^3$He atom
on a Cs substrate is also given.
The dotted line has been drawn as a to guide the eye.}
\label{fig2}
\end{figure}

\begin{figure}
\caption[]
{Spectrum of a single $^3$He atom as a function of squared angular 
momentum for the bands $n$ = 1 to 4, together with the KS energies of the occupied sp  
levels corresponding  to $N_3=26$ (open circles), with dotted lines 
to guide the eye.}
\label{fig3}
\end{figure}

\begin{figure}
\caption[]
{The EL/KS $^3$He  and $^4$He  particle densities
for $N_3=26$ and $N_4=1000$.}
\label{fig4}
\end{figure}

\begin{figure}
\caption[]
{Threedimensional view of the $^3$He particle density in the $y=0$ plane
for $N_4=1000$ and
 $n_3 = 3 \times 10^{-3} {\rm \AA}^{-2}$.
The highest density peaks correspond to $\rho_3 \sim 1.4 \times 10^{-2}
\,{\rm \AA}^{-3}.$
}
\label{fig5}
\end{figure}

\begin{figure}
\caption[]
{Threedimensional view of the $^4$He particle density in the $y=0$ plane
for $N_4=1000$
and  $n_3 = 3 \times 10^{-3} {\rm \AA}^{-2}$.
The highest density ridge corresponds to $\rho_4 \sim 2.4 \times 10^{-2}
\,{\rm \AA}^{-3}.$
}
\label{fig6}
\end{figure}

\begin{figure}
\caption[]
{Contour plots of the total $^3$He$+^4$He density
corresponding to 0.011 ${\rm \AA}^{-3}$ for the case  $N_4=1000$
and  $n_3 = 2 \times 10^{-3} {\rm \AA}^{-2}$,
$n_3 = 3 \times 10^{-3} {\rm \AA}^{-2}$. The dashed line corresponds
to the pure $^4$He drop.}
\label{fig7}
\end{figure}

\end{document}